\documentclass[iop]{emulateapj}

\newcommand{\degrees}{\mbox{$^{\circ}$}} 
\def\sqig{$\sim$}

\def\ls5039{LS\,5039}
\def\J1018{1FGL\,J1018.6$-$5856}
\def\j1018{1FGL\,J1018.6$-$5856}

\def\msun{$M_{\odot}$}
\def\rsun{$R_{\odot}$}

\def\ergcm2s{erg\,cm$^{-2}$\,s$^{-1}$}
\def\ergscm2s{erg\,cm$^{-2}$\,s$^{-1}$}
\def\mevcm2s{MeV\,cm$^{-2}$\,s$^{-1}$}
\def\MeVcm2s{MeV\,cm$^{-2}$\,s$^{-1}$}
\def\phcm2s{ph\,cm$^{-2}$\,s$^{-1}$}
\def\pphcm2s{p\,ph\,cm$^{-2}$\,s$^{-1}$}

\def\Fermi{{Fermi}}

\def\src{4FGL\,J1702.7$-$5655}

\def\bfa{}
\def\bfb{}

\def\mybf{}

\usepackage{draftcopy}
\usepackage[utf8x]{inputenc}
\usepackage{amsmath}

\draftcopyName{}{24}

\begin{document}
\def\subtitle{}
\submitted{}
\accepted{May 13, 2022}
\journalinfo{}

\title{Gamma-ray Eclipses and Orbital Modulation Transitions in the Candidate Redback 4FGL J1702.7-5655}

\author{R.~H.~D. Corbet\altaffilmark{1,2}, L. Chomiuk\altaffilmark{3}, 
J.~B. Coley\altaffilmark{4}, G. Dubus\altaffilmark{5}, \\
P.~G. Edwards\altaffilmark{6}, N. Islam\altaffilmark{1}, V.~A. McBride\altaffilmark{7}, 
J. Stevens\altaffilmark{6}, J. Strader\altaffilmark{3}, 
S.~J. Swihart\altaffilmark{8},
L.~J. Townsend\altaffilmark{9}
}

\altaffiltext{1} {University of Maryland, Baltimore County, and
X-ray Astrophysics Laboratory, Code 662 NASA Goddard Space Flight Center, Greenbelt Rd., MD 20771, USA.}

\altaffiltext{2} {
Maryland Institute College of Art, 1300 W Mt Royal Ave, Baltimore, MD 21217, USA.}

\altaffiltext{3} {Department of Physics and Astronomy, Michigan State University, East Lansing, MI 48824, USA.}

\altaffiltext{4} {Department of Physics and Astronomy, Howard University, Washington, DC 20059, USA. 
CRESST/Code 661 Astroparticle Physics Laboratory, NASA Goddard Space Flight Center, Greenbelt Rd., MD 20771, USA.}

\altaffiltext{5} {Institut de Plan\'{e}tologie et d'Astrophysique de Grenoble, Univ. Grenoble Alpes, CNRS, F-38000 Grenoble, France.}

\altaffiltext{6} {Commonwealth Scientific and Industrial Research Organisation Astronomy and Space Science, PO Box 76, Epping, New South Wales 1710, Australia.}

\altaffiltext{7} {Department of Astronomy, University of Cape Town, Private Bag X3, Rondebosch, 7701, South Africa.}

\altaffiltext{8} {National Research Council Research Associate, National Academy of Sciences, Washington, DC 20001, USA,\\ resident at Naval Research Laboratory, Washington, DC 20375, USA.}

\altaffiltext{9} {South African Astronomical Observatory, PO Box 9, Observatory, 7935, South Africa.}

\begin{abstract}
Observations with
the \Fermi\ Large Area Telescope (LAT) of the gamma-ray source \src, previously classified as a candidate
millisecond pulsar, show
highly-significant modulation at a period
of 0.2438033 days (\sqig 5.85 hours).
Further examination of the folded light curve indicates the presence of narrow eclipses, suggesting this is a redback binary system.
An examination of the long-term properties of the modulation over
13 years of LAT observations indicates that the orbital modulation of
the gamma-rays changed from a simple eclipse before early 2013, to a broader, 
more easily detected,
quasi-sinusoidal modulation.
In addition, the time of the eclipse shifts to \sqig0.05 later in phase.
This change in the orbital modulation properties is, however, not accompanied by a significant overall change in gamma-ray flux
or spectrum.
The quasi-sinusoidal component peaks \sqig0.5 out of phase with the eclipse, which would
indicate inferior conjunction of the compact object in the system. 
Swift X-ray Telescope observations reveal a possible X-ray counterpart within the {\mybf LAT} error ellipse.
However, radio observations obtained with the Australia Telescope Compact Array do
not detect a source in the region.
\src\ appears to have changed its state in 2013, perhaps related to changes in the
intrabinary shock in the system.
We discuss how the properties of \src\ compare to other binary millisecond pulsars that have
exhibited orbital modulation in gamma rays.

\end{abstract}

\section{Introduction}
\subsection{Binary Millisecond Pulsars}

Millisecond pulsars (MSPs) are pulsars with short pulse periods, but which are apparently
old as indicated by their slow spin-down rates. They are believed to be descended from low-mass X-ray binaries (LMXBs).
Some MSPs are found in ``spider'' binaries with a low-mass companion that
is being ablated by the wind from the pulsar \citep[e.g.][]{Roberts2013}. The spider binaries are generally divided into black widow systems,
in which the companion is very low mass and may be degenerate, and redbacks where the companion is
a near main-sequence star. 
An intrabinary shock (IBS) is thought to exist between the winds from the pulsar and its
companion, and this can be the site of X-ray and radio emission \citep[e.g.][]{Gentile2014,Roberts2014,Kandel2019,vanderMerwe2020}.
Orbitally modulated emission is commonly seen both in X-rays, associated with accelerated particles
in the IBS, and at radio wavelengths where eclipses can be caused by scattering
in dense ionized material.
The optical emission is dominated by the light from the pulsar's companion, and this is
generally modulated on the orbital period due to both gravitational ellipsoidal distortion
and heating of the face of the companion that is facing the pulsar \citep[e.g.][]{Breton2013}. 

Several redbacks have been seen to change between
states in which the primary power source is the rotational energy of the neutron star,
and those in which
accretion is occurring from the companion, and it becomes
an LMXB. These are termed ``transitional millisecond pulsars''  \citep[tMSPs, e.g.][]{Papitto2020}.
In tMSPs three states are generally identified:
(i) a pulsar state where the pulsar is the power source in the system
and accretion is not occurring; 
(ii) accretion states where accretion onto the surface of the neutron star occurs;
(iii) a sub-luminous disk state where there can be rapid changes on timescales
as short as \sqig10 s between
high and low X-ray flux levels, with intermittent flares also present.
Gamma-ray brightening may accompany a change from the rotation-powered state
to the sub-luminous state.

The Large Area Telescope (LAT) on board the Fermi {\mybf Gamma-ray Space Telescope} has been surveying the gamma-ray
sky above 100 MeV since 2008, and it has now identified more than 5000 sources.
Of these, a significant number \citep[$>$ 260; e.g.][]{Griessmeier2021} 
have been identified as pulsars, with a large fraction being MSPs\footnote{https://confluence.slac.stanford.edu/display/GLAMCOG/Public+List+of+LAT-Detected+Gamma-Ray+Pulsars}.
Detection of periodic orbital modulation in MSPs at gamma-ray energies is relatively uncommon, compared to that at other energies.
4FGL\,J0427.8$-$6704 (3FGL\,J0427.9$-$6704)
was previously found to exhibit eclipses in gamma rays, X-rays, and the optical \citep{Strader2016,Kennedy2020} 
at a period of \sqig0.366 days.
More recently, \citet{Clark2021b} found {\mybf what they describe as} ``subtle'' gamma-ray eclipses
in four systems: 4FGL\,J1048.6$+$2340 (PSR J1048+2339), 4FGL\,J1816.5$+$4510 (PSR J1816+4510),
4FGL\,J2129.8$-$0428 (PSR J2129-0429), and PSR B1957+20.
The detection of these eclipses exploited orbital ephemerides obtained from radio and
gamma-ray pulse timing, which give both the orbital period
and the epoch of expected eclipse.

We have been conducting a search for previously unrecognized gamma-ray emitting
binaries in the Fermi{\mybf -LAT}  catalogs
by searching for periodic modulation of the LAT flux. This has enabled us
to detect several high-mass systems \citep[][and references therein]{Corbet2019}. 
Here we present the detection of strong periodic gamma-ray modulation in LAT observations of the candidate redback system \src.
We find that it exhibits periodic dips in its light curve that are consistent with 
eclipses in a redback system.
In addition, an investigation of the multi-year 
behavior indicate a change in modulation properties from just a periodic dip to quasi-sinusoidal
modulation together with a dip at a slightly later phase.
We also searched for X-ray and radio counterparts of \src\ using the Swift X-ray Telescope (XRT) 
and the Australia Telescope Compact Array (ATCA) respectively, and identify a possible
X-ray counterpart although no counterpart is detected at radio wavelengths.
We compare the gamma-ray modulation in \src\ to that in other systems, and speculate
on the cause of the change in orbital modulation.
Unless otherwise noted, uncertainties are given at the 1$\sigma$ level.

\subsection{Previous Observations of \src\label{sect:source}}

\src\ is in the fourth LAT catalog \citep{Abdollahi2020} and
counterparts were also present in the third catalog \citep[3FGL\,J1702.8$-$5656;][]{Acero2015},
the LAT eight year source list\footnote{https://fermi.gsfc.nasa.gov/ssc/data/access/lat/fl8y/} (FL8Y\,J1702.7-5654),
the second catalog \citep[2FGL\,J1702.5-5654;][]{Nolan2012},
and the first catalog \citep[1FGL\,J1702.4-5653;][]{Abdo2010}.
It may also be associated with the AGILE gamma-ray source 2AGL\,J1703$-$5705 \citep{Bulgarelli2019}.
\citet{Saz2016} undertook a classification of sources in
the 3FGL catalog into pulsars and active galactic nuclei - the two main categories
of identified LAT sources. From this analysis they found that 
3FGL\,J1702.8$-$5656 was most likely to be a millisecond pulsar.
3FGL\,J1702.8$-$5656 was included in a search for gamma-ray pulsations from Fermi LAT observations
for frequencies $<$ 1520 Hz by 
\citet{Clark2017}  and \citet{Wu2018}, but none were reported.
{\mybf In addition, 3FGL\,J1702.8$-$5656 was included in a search for radio
pulsations using the Parkes telescope by \citet{Camilo2015} made with 125$\mu$s time resolution
and none were found from this source.}

\section{Observations and Analysis}

\subsection{Gamma-ray Observations and Analysis\label{sect:gray_obs}}

The \Fermi\ LAT \citep{Atwood2009,Ajello2021} is a pair conversion telescope 
sensitive to gamma-ray photons with energies
between \sqig20 MeV to $>$ 300 GeV.
The LAT data used in this paper were obtained between 2008 August 4 and 2021 August 19  (MJD 54,682
to 59,445). 
During this time, Fermi was primarily operated in a sky survey mode.
Until 2018 
the LAT pointing position was alternately rocked away from the zenith to the orbit
north for one spacecraft orbit, then towards the orbit south for
one orbit. In this way, the entire sky was observed every two spacecraft orbits, approximately
every three hours. After a failure of the drive for one solar array, changes were made to the sky survey profiles, 
but coverage of the entire sky was maintained in the long term \citep{Ajello2021}.

Our search for new gamma-ray binaries followed similar procedures to those described in \citet{Corbet2019}. For LAT analysis we used the \texttt{fermitools} version \texttt{1.0.1} with
the updated Pass 8 LAT data files \citep[``P8R3'',][]{Bruel2018} and the weekly photon files 
provided by the Fermi Science Support Center which include precomputed diffuse response columns.
Light curves covering an energy range of 100 MeV to 500 GeV were
created for all 4FGL DR2 sources using time bins of 500 s. Times
when sources were closer than 5 degrees to the Sun were excluded, but no filtering was
applied for distance from the Moon.
The light curves were obtained using a variant of aperture photometry
where we estimate the probability ``$p$'' that each photon comes from
a source of interest and sum these probabilities \citep[e.g.][]{Kerr2011,FermiLAT2012,Kerr2019}. 
{\mybf To facilitate this, model files were created for each source using \texttt{make4FGLxml} including
sources from the 4FGL DR2 catalog 
within a 10 degree radius from the source.} 
Photon probabilities were calculated using \texttt{gtsrcprob} and
then summed for a 3 degree radius aperture centered on each source.
{\mybf We used \texttt{gtbin} to create the initial light curves, but
then replaced the integer \texttt{COUNTS} column with a floating
point \texttt{RCOUNTS} column containing the calculated summed probabilities.}
The exposure for each time bin was calculated using \texttt{gtexposure}
to obtain the probability weighted rate in units of \pphcm2s\
and times were corrected to the Solar System barycenter with \texttt{gtexposure}.
We note that although the use of ``probability photometry'' generally
increases the signal-to-noise of the light curves, it affects the
photometric properties as probabilities are based on a constant
source brightness. Thus, when a source is brighter than the model
predicts, the probability of a photon coming from the source is underestimated,
and, when the source is fainter than the model prediction, the probability is overestimated
\citep[e.g.][]{Kerr2019}. 
This results in a reduction of the apparent amplitude of any variability.

Power spectra of these LAT light curves were calculated {\mybf using Discrete Fourier Transforms (DFT)}, weighting each
data point's contribution by its relative exposure after first subtracting
the mean count rate. This is beneficial because of the substantial exposure changes
from time bin to time bin. 
{\mybf We note that while the time bins produced by \texttt{gtbin} are evenly
spaced, although with gaps, this is no longer the case after the barycenter correction. However, even spacing is not required for calculation of a DFT.}
{\mybf The use of weighting does not, however, change the intrinsic statistical properties
of the power spectrum \citep{Zechmeister2009,VanderPlas2018} and the powers are
expected to be distributed given by an exponential probability function.
We normalize the computed powers to have a average power of unity.
}
The power spectra cover a period range from 0.05 days
(1.2 hrs) to the length of the light curve, i.e. \sqig 4762 days, giving \sqig95,246 independent
frequencies. {\bfa This is the same period range that we have previously used in our searches \citep{Corbet2016,Corbet2019}.} {\bfb While this range was primarily chosen to 
facilitate a search for high-mass binaries, encompassing the relatively short 0.2 day period of Cygnus X-3 (4FGL\,J2032.6+4053) for which modulation of the LAT flux is already known \citep[e.g.][]{Corbel2012}, 
we can potentially detect periodic modulation from any
type of source in this range.}
The power spectra were oversampled by a factor of 5 compared to the nominal
resolution, which we take to be the inverse of the length of the
light curve {\mybf(``$T$'')} \citep[e.g.][and references therein]{VanderPlas2018}, i.e. \sqig1/4762 days$^{-1}$.
{\bfb As has been noted by several authors, if the power spectrum is not oversampled
then the measured power of a signal will be lower than the true power by a factor that
ranges between \sqig 0.405 and 1, with a mean of \sqig0.773 
\citep[e.g.][]{vanderKlis1989,Vaughan2005,VanderPlas2018}. 
The lowest measured
power will occur if the signal frequency lies halfway between the sampling intervals.
We also note that the same effect will also apply to noise peaks in the power spectrum.
} 
For the strongest peak in each power spectrum
the False Alarm Probability \citep[FAP,][]{Scargle1982}, the estimated probability 
of a signal reaching a power level by chance assuming white noise,
was calculated.
{\mybf 
\begin{equation}
FAP = 1 - [1 - \exp(-r)]^N
\end{equation}
where $r$ is the normalized height of the peak power, and $N$ is the number of independent frequencies.
}
{\bfa As noted by \citet{Scargle1982} the oversampling is effectively interpolation and so does not increase
the number of independent frequencies.}
This {\mybf calculation of the} FAP takes into account the number of independent frequencies
searched, {\mybf correcting for the oversampling of the power spectra,} but does not include the effect of searching for periodicity in multiple
sources. 
{\mybf In the case where we are examining a power spectrum for the presence of a previously reported periodicity with a small uncertainty, $N$ may be
set equal to unity, and hence a lower amplitude signal can give a smaller FAP than the case for a ``blind'' search.
}
{\mybf
The uncertainty in the frequency of a candidate modulation that is detected can be calculated using the derivation of \citet{Horne1986} and as 
also discussed by \citet{Levine2011}:
\begin{equation}
\delta f = \frac{3}{8} \frac{1}{T\sqrt{r}}
\end{equation}

Since the period, $P$, = 1/$f$, propagation of errors yields:
\begin{equation}
\begin{aligned}
\frac{\delta f}{f} = \frac{\delta P}{P} \\
\Rightarrow \delta P = \frac{3}{8} \frac{P^2}{T\sqrt{r}} \\
\end{aligned}
\end{equation}
Hence, the fractional uncertainty of the frequency and period 
\begin{equation}
\frac{\delta f}{f} = \frac{\delta P}{P} \propto f^{-1}
\end{equation}
i.e. the fractional uncertainty on the measurement of the frequency and period of a detected
modulation is smaller for short periods/high frequencies than long periods/low frequencies.
In addition, the degree to which a frequency can be measured more precisely than
the intrinsic Fourier resolution, $\pm 1/(2T)$, scales as the square root of the normalized
peak height.
}
{\bfa Hence, after a significant peak in the power spectrum has been detected, to determine
the frequency to better than the uncertainty implied by Eqn. 2, it is necessary to calculate
the power spectrum in the vicinity of the peak with an oversampling ratio significantly $> 8/3 \sqrt{r}$.
We typically choose to oversample by a factor of 20 greater than this, i.e. $\sim 50 \sqrt{r}$.}

In our photometric analyses the background is not fitted for each time bin, and artifact signals can be
seen at several periods including \Fermi's \sqig 90 minute orbital period, the survey period at twice this,
one day, the Moon's 27.3 day sidereal period,
the 53 day precession period of the \Fermi\ satellite's orbit, and one quarter of a year related to the shape of the LAT PSF\footnote{http://fermi.gsfc.nasa.gov/ssc/data/analysis/LAT\_caveats\_temporal.html}.

\subsection{X-ray Observations and Analysis\label{sect:xray_obs}}

The Swift-XRT \citep{Burrows2005} is a Wolter I X-ray imaging telescope sensitive to X-rays ranging 
from 0.3 to 10 keV. 
The location of \src\ had previously been observed for 4ks with the XRT 
under the program of \citet{Stroh2013} to monitor unassociated LAT sources. 
As no source was clearly detected in these observations, additional observations
were obtained under the Swift TOO program to acquire a total of \sqig 17.5 ks of exposure.
All observations were made in photon counting (PC) mode.
The log of XRT observations is given in Table \ref{table:xrt}.
We analyzed the existing and new data sets together using the online tools
provided at the University of Leicester \citep{Goad2007,Evans2009} to extract an image,
and for source detection and position determination\footnote{https://www.swift.ac.uk/user\_objects/}.

\subsection{Radio Observations and Analysis\label{sect:radio_obs}}

Radio observations covering the location of \src\ were obtained using ATCA \citep{Wilson2011} 
between December 2020 to November 2021
(MJD 59208 to 59540, see Table \ref{table:atca}), i.e. during Part 2 of the gamma-ray light curve
(Section \ref{sect:gray_results}), with observations centered at 2.1, 5.5 and 9.0 GHz, 
with 2 GHz bandwidths for all three bands. The ATCA, which consists of six 22 m-diameter antennas, 
was in several different array configurations over this period, with the more compact arrays
somewhat more sensitive to the bright extended emission in the vicinity. Details of
the array configurations are given in Table \ref{table:atca}.
Observations were reduced following standard procedures in Miriad \citep{Sault1995}. 
PKS 1657-56 was used as phase calibrator and
PKS 1934-638 was used as a primary flux density calibrator for all observations.

\section{Results\label{sect:results}}

\subsection{Gamma-ray Results\label{sect:gray_results}}

\begin{figure}
\includegraphics[width=7.25cm,angle=270]{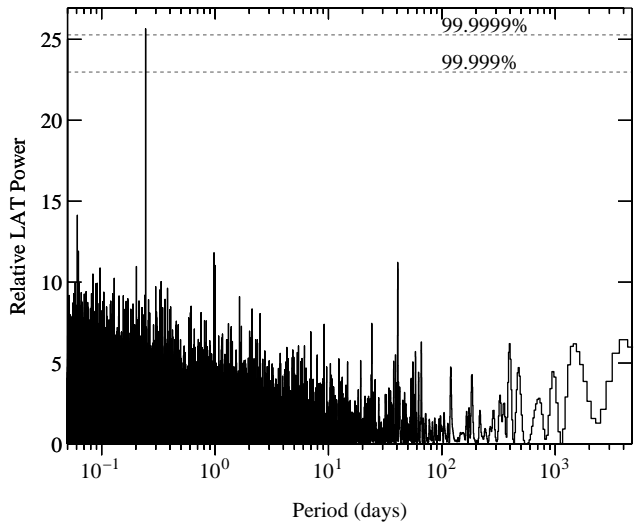}
\caption{
Power spectrum of the \Fermi\ LAT probability-weighted
aperture photometry light curve of \src.
{\bfa The power spectrum is normalized to the mean power level
of 1.3\,$\times$\,10$^{-19}$ (\pphcm2s)$^{2}$.}
}
\label{fig:lat_power}
\end{figure}

From our examination of the power spectra of the LAT light curves of all 4FGL DR2 sources we noted
significant modulation from \src.
The power spectrum of the LAT light curve is shown in Fig. \ref{fig:lat_power}.
There is a prominent peak near a period of 0.24 days at a height of
25.7 times the mean power level {\bfa of 1.3\,$\times$\,10$^{-19}$ (\pphcm2s)$^{2}$} and the {\bfa implied} FAP is 7$\times$10$^{-7}$.
{\mybf To investigate the frequency dependence of the continuum power 
we performed a fit of the logarithm of the power as a function of the
logarithm of the frequency as advocated by \citet{Vaughan2005}. From this we find
that the continuum is very flat with only a hint of a small increase at lower
frequencies/longer periods with 
$Power \propto f^{-4(\pm 2) \times 10^{-3}}$.
{\bfa We also calculated the mean power level around the peak in the period range
0.24 to 0.25 days and the peak has a relative height of 24.5, which gives an FAP of 2.2$\times$10$^{-6}$,
calculated using the total number of independent frequencies for the full period range.} 
We note that the apparent
increase in power at shorter periods in Fig. \ref{fig:lat_power} is not real and
is due to the logarithmic scale which prevents individual points being discerned
at high frequencies,f and only the statistical envelope being visible.
}

\begin{figure}
\includegraphics[width=7.25cm,angle=270]{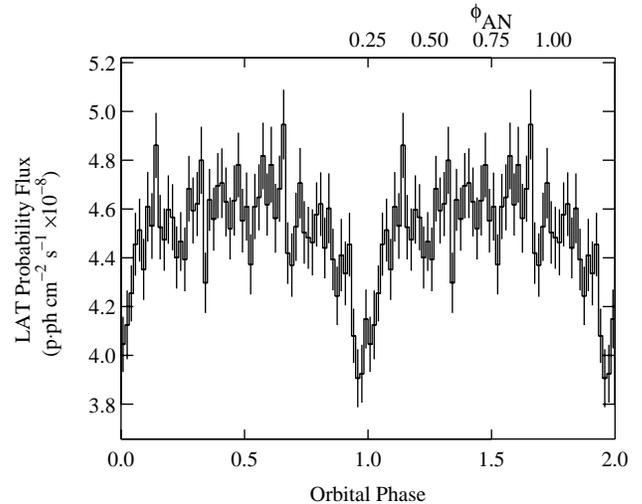}
\caption{
Fermi LAT probability-weighted aperture photometry light curve of \src\ 
folded on the proposed orbital period. The bottom X-axis uses
eclipse center to define orbital phase zero. The upper labels
show the predicted orbital phases with time of ascending node defining zero.
}
\label{fig:lat_fold}
\end{figure}

No obvious peaks are seen at the second, third, or fourth harmonics
of this. 
The period is determined to be 0.2438033(11) days using the formulation of \citet{Horne1986},
i.e. 5.851279(26) hours.
Since this would be a typical orbital period for a binary MSP \citep[e.g.][]{Papitto2020}, 
we assume that this is indeed the orbital period of \src.
The probability-weighted aperture photometry LAT light curve folded on the 0.24 day period is shown in Fig. \ref{fig:lat_fold}.
This shows a relatively sharp dip 
that is strongly suggestive of the presence of an eclipse, implying that the system 
is observed at a high inclination angle.

\begin{figure}
\includegraphics[width=8.5cm,angle=0]{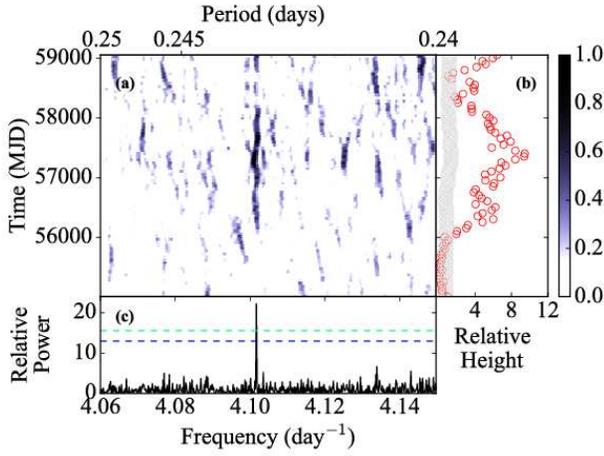}
\caption{
{\mybf
(a) Dynamic power spectrum of the probability-weighted
aperture photometry LAT light curve of \src.
(b) Relative peak at the orbital period to the mean power of
the values shown in panel (a).
(c) Coherent power spectrum of the entire light curve.
}
}
\label{fig:lat_dynamic}
\end{figure}

\begin{figure}
\includegraphics[width=7.25cm,angle=270]{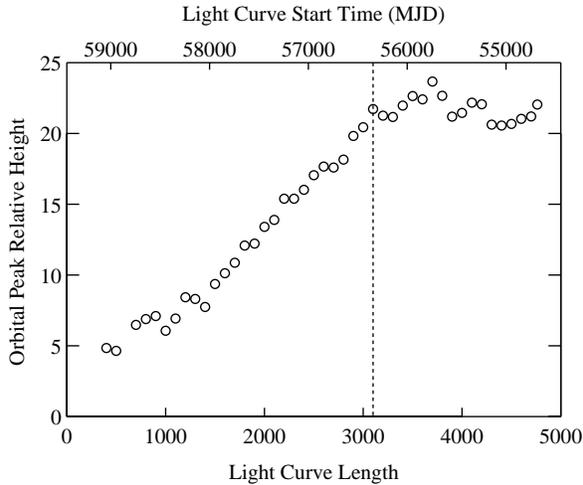}
\caption{
Relative height of the orbital peak in the power spectrum
of the probability-weighted aperture photometry LAT light curve of \src. All light curves have the same end date of MJD 54,682, and
the start date, shown in the top axis, is changed in the analysis.
}
\label{fig:lat_rvmt}
\end{figure}

In order to search for any long-term changes in the orbital modulation
we obtained a dynamic power spectrum of the probability-weighted aperture
photometry LAT light curve.
We calculated power spectra for light curves of length 650 days, with offsets
of 50 days between successive light curves. The results of this
are shown in Fig. \ref{fig:lat_dynamic} and it
suggests that the orbital modulation was less apparent during
earlier time ranges.
To explore this further, we investigated the growth in relative height
of the peak in the power spectrum as a function of time.
For a persistent coherent signal with constant background and no other changes, the relative
power should grow linearly with time as
the amplitude noise decreases as the square root of the observation time.
In Fig. \ref{fig:lat_rvmt} we plot the relative height of the orbital peak
using light curves that all end at the same time (MJD 59,445) but have different start dates.
From this, it can be seen that as the start date is moved earlier, the relative strength of
the peak initially grows as the light curve length increases. 
However, for light curve start times earlier than \sqig3100 days before
the end of the light curve (i.e. start times earlier than \sqig MJD 56,345 = 2013-02-22)
a plateau appears with little overall increase in relative power.

\begin{figure}
\includegraphics[width=7.25cm,angle=0]{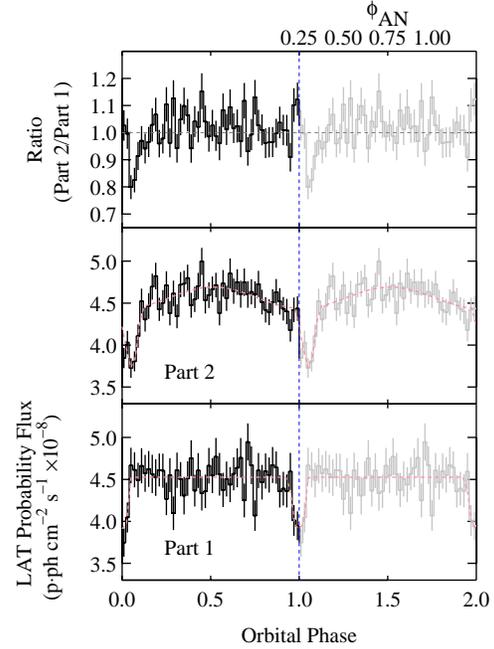}
\caption{
The probability-weighted
aperture photometry LAT light curve of \src\ obtained
for time ranges of MJD 54,682 - 56,345 (bottom) and 56,345 - 59,445 (middle)
folded on the orbital period. The top panel shows the ratio of the count rate
in the second part to the first part. The dashed pink lines in the bottom two
panels indicate the fits made to the unfolded light curves, the parameters
are given in Table \ref{table:lc_fits}.
For clarity, for all panels two identical cycles are plotted with the second
cycle plotted in gray.
The bottom X-axis uses
eclipse center to define orbital phase zero. The upper labels
show the predicted orbital phases with time of ascending node defining zero.
}
\label{fig:lat_two_fold}
\end{figure}

\begin{figure}
\includegraphics[width=7.25cm,angle=0]{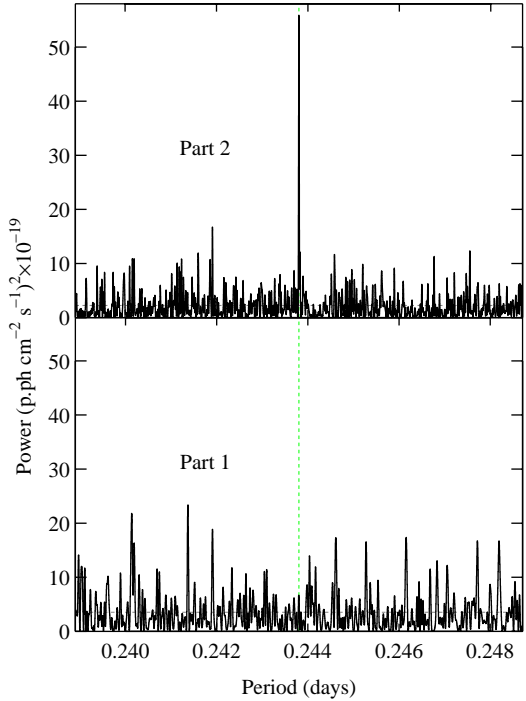}
\caption{
Power spectra of the probability-weighted
aperture photometry LAT light curve of \src\ obtained
for time ranges of MJD 54,682 - 56,345 (bottom) and 56,345 - 59,445 (top).
The period ranges are centered on the presumed orbital period marked with
the vertical dashed green line. 
The mean power level for each panel is marked with a dashed horizontal gray line.
Note that the y axis is in units of absolute, rather than
relative, power and that the same range is plotted for both panels.
}
\label{fig:lat_two_power}
\end{figure}

To investigate this in more detail we calculated power spectra for the LAT
light curve divided into two sections before and after MJD 56,345. 
These are shown in Fig. \ref{fig:lat_two_power} and it can be seen
that there is no significant orbital peak in the power spectrum for the earlier
light curve segment (``Part 1'') while it is strongly detected for the later segment (``Part 2''). 
We then folded the light curve on the orbital
period using data from these two time ranges and these are shown in
Fig. \ref{fig:lat_two_fold}. From these it can be seen that the time of the minimum
is shifted to the right and is broader for Part 2 compared to Part 1.
In Part 2 the weighted count rate outside the eclipse
is also slightly higher as indicated by the upper panel of Fig. \ref{fig:lat_two_fold}
which shows the ratio of the two folded light curves.

\clearpage

We note that we cannot determine the exact time of the change in orbital
modulation due to the long integration times required to detect the
modulation and changes in it. We also investigated the effect on the power spectrum
dividing the light curve into halves with a 600 day earlier split (MJD 55,745) where
the relative peak in Fig. \ref{fig:lat_rvmt} is at its maximum. In this case,
we find that the {\em absolute} power for Part 2 is reduced by \sqig12\% compared to dividing
the light curve at MJD 56,345 (Fig. \ref{fig:lat_two_power}). 
Thus, any sinusoidal orbital modulation between MJD 55,745 to 56,345 would be at a lower amplitude.

\begin{figure}
\includegraphics[width=7.25cm,angle=0]{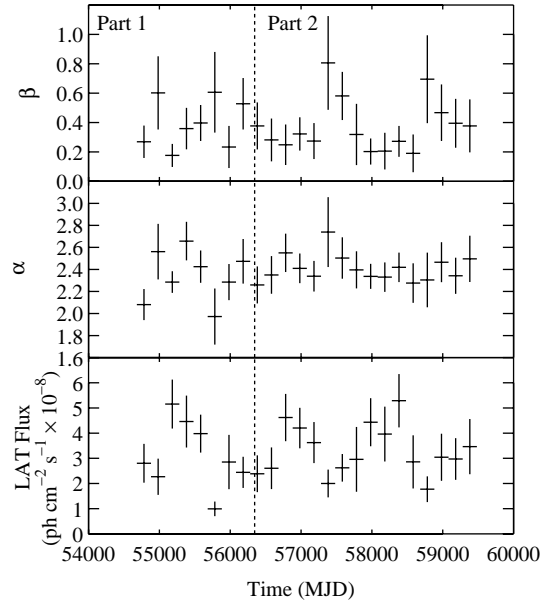}
\caption{
Long-term Fermi LAT light curve of \src\ obtained from a maximum likelihood analysis.
{\mybf The energy range is 100 MeV to 500 GeV and the}
spectral parameters are for the logParabola model
used in the 4FGL catalog.
The dashed line shows the division between Part 1 and Part 2 at MJD 56,345.
}
\label{fig:lat_spec_lc}
\end{figure}

In order to search for other long-term changes in the gamma-ray properties that, for example,
could be an indication of a state change in a tMSP we calculated a light
curve using {\mybf a binned} likelihood analysis.
We performed likelihood fitting 
using time bins of 200 days, 
{\mybf an energy range of 100 MeV to 500 GeV,}
a region of interest of 10 degrees,
and the spectral parameters for \src\ were allowed to float, while spectral values for other
sources in the region were held fixed. 
The spectral model used for \src\ in the 4FGL catalog is a log normal function (\texttt{LogParabola}), 
i.e.
\begin{equation}
\frac{dN}{dE} = K \left(\frac{E}{E_0}\right)^{-\alpha-\beta\,log_e(E/E_0)}
\end{equation}
This model is used in the LAT catalogs for all sources with significantly
curved spectra. 
The resulting light
curve is shown in Fig. \ref{fig:lat_spec_lc} and the approximate time where
the orbital modulation changed to become prominent in the light
curve is marked. Although the light curve is formally inconsistent
with the hypothesis of being constant ($\chi^2_\nu$ = 3.1), we do not see any clear changes in the flux
level or spectral parameters associated with the change in the orbital
profile. The mean flux before the division is 2.5 $\pm$ 0.3 $\times$10$^{-8}$\,\phcm2s\ 
{\mybf (1.8  $\pm$ 0.2 $\times$10$^{-11}$\,\ergscm2s)},
while after it is 2.9 $\pm$ 0.3 $\times$10$^{-8}$\,\phcm2s\ {\mybf (2.1 $\pm$ 0.2 $\times$10$^{-11}$\,\ergscm2s)}. 
For $\alpha$, the values before and after are 2.36 $\pm$ 0.03 and 2.38 $\pm$ 0.3, while for
$\beta$ they are 0.30 $\pm$ 0.05 and 0.30 $\pm$ 0.04, respectively.

In order to characterize the orbital modulation we fitted periodic functions to
the {\em unfolded} weighted aperture photometry light curves.
Since the eclipse has a relatively short duration, we extracted light curves
with 100 s time bins, i.e. \sqig 0.005 of the orbital period.
We fitted the two parts of the light curve separately.
For Part 1 we used an eclipse profile with constant fluxes outside and inside
the eclipse, and linear transitions between these, with independent durations
for the eclipse ingress and egress.
The profile is defined in terms of time of orbital period {\bfb (which is held fixed
at the value found from the power spectrum)} and eclipse center {\bfb (T$_0$)}, which we 
define as $\phi$ = 0, the phase of the start of eclipse ingress ($\phi_{\rm ing}$),
the phase of the end of egress ($\phi_{\rm egr}$), the total duration of the eclipse minimum ($\Delta$),
flux outside eclipse ($F_{\rm unecl}$), and flux during eclipse totality ($F_{\rm ecl}$).
In the fits $\phi_{\rm ing}$ and $\phi_{\rm egr}$ were constrained to occur before and after
the phases of eclipse totality respectively.
Similarly to the computation of the power spectra, the data points were weighted by their relative exposures.

\begin{equation}
\label{Flux Equation}
  F =
  \left\{
    \begin{array}{ll}
      F_{\rm unecl}, 
& \phi_{\rm egr} \leq \phi \leq \phi_{\rm ing} \\
      F_{\rm unecl} + (\frac {(F_{\rm ecl} - F_{\rm unecl}) (\phi -  \phi_{\rm ing})}{(1 - (\frac {\Delta}{2}) - \phi_{\rm ing})})   , 
& \phi_{\rm ing} \leq \phi \leq (1 - (\frac {\Delta}{2})) \\
      F_{\rm ecl}, 
& \phi \ge (1 - (\frac{\Delta}{2})); \phi \leq \Delta/2\\
	F_{\rm ecl} + (\frac{(F_{\rm unecl} - F_{\rm ecl})(\phi - (\frac {\Delta}{2})}{(\phi_{\rm egr} - (\frac {\Delta}{2}))}), 
&  (\frac {\Delta}{2}) \leq \phi \leq \phi_{\rm egr} \\

    \end{array}
  \right.
\end{equation}

Derived from these fitted parameters are the total eclipse phase duration from eclipse ingress
start to egress end (1 + $\phi_{\rm egr}$ - $\phi_{\rm ing}$), the phase duration of the ingress
(1 - ($\frac{\Delta}{2}$) - $\phi_{\rm ing}$), 
and the phase duration of the egress ($\phi_{\rm egr}$ - ($\frac {\Delta}{2}$)).

For the fit to the unfolded Part 2 light curve we used a sum of the same eclipse model together with an additional sine wave component.
For both fits we held the orbital period at the value determined from the power spectrum of Part 2.
If we allowed the orbital period to be free we found that this resulted in unstable fits.

The parameters from the fits to both sections of the light curve are given in Table \ref{table:lc_fits}. {\bfa The parameter 1 $\sigma$ uncertainties are derived from the intervals that give $\chi^{2}_{min} + 1$ \citep{Lampton1976,Yaqoob1998}.}

We use the center of the eclipse for the Part 1 light curve 
to define phase zero and find this to be MJD 57000.168, with
an uncertainty of 0.004 days (uncertainty of 0.017 in phase).
This would correspond to superior conjunction of a pulsar, assuming that 
is indeed the source of the gamma-rays and it is being eclipsed by
its companion. For a pulse timing circular orbit with phase 0 defined as the ascending node, this would correspond to
a phase ($\phi_{AN}$) of 0.25. 
The Part 1 eclipse width from ingress start to egress end is 0.110 $\pm$ 0.038 in
phase. The ingress and egress durations are not well defined and are consistent with 0.

For the Part 2 light curve, the center of the eclipse minimum is shifted to a slightly later time
by 0.014 $\pm$ 0.005 days. i.e. our fits indicate that the centers of the two eclipses differ 0.059 $\pm$ 0.020 in phase.
The semi-amplitude of the sine wave component is 0.127 $\pm$ 0.034 $\times$10$^{-8}$ \pphcm2s, {\mybf equivalent to a fractional semi-amplitude of 2.8 $\pm$ 0.7 \%.}
This corresponds to a power of 16 $\pm$ 9 $\times$10$^{-19}$\,(p.ph\,cm$^{-2}$\,s$^{-1}$)$^2$.
The measured power for the orbital modulation in Part 2 is \sqig55 $\times$10$^{-19}$\,(p.ph\,cm$^{-2}$\,s$^{-1}$)$^2$ 
(Fig. \ref{fig:lat_two_power})
and so, at least in this representation of the light curve components, the eclipse feature
also contributes to the observed signal in the power spectrum. This may be due to the eclipse,
which occurs at the minimum of the sine component, becoming broader.
We note, however, that this mathematical deconvolution does not necessarily relate to two astrophysically distinct
components.

\begin{figure}
\includegraphics[width=7.25cm,angle=0]{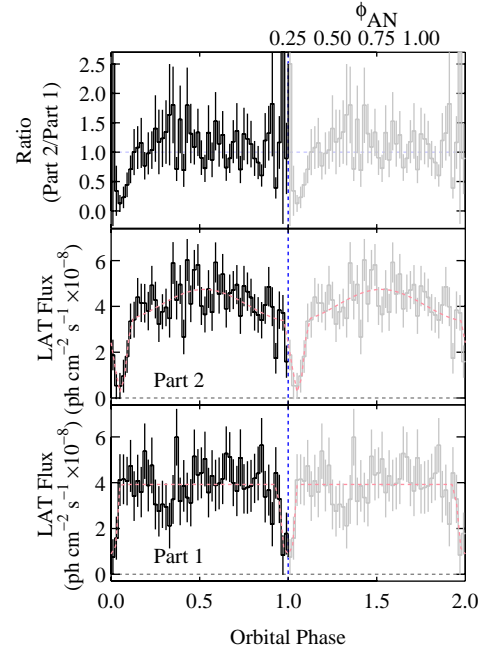}
\caption{
The binary phase-resolved flux of \src\ obtained
from likelihood analysis of LAT observations. Spectral parameters were held fixed at the 4FGL catalog (DR2) values.
Fits are shown for time ranges of MJD 54,682 - 56,345 (bottom) and 56,345 - 59,445 (middle).
The top panel shows the ratio of the fluxes in the second part to the first part.
The dashed pink lines in the bottom two
panels indicate the fits made to the folded light curves, the parameters
are given in Table \ref{table:like_fold_fits}.
For clarity, for all panels two identical cycles are plotted with the second
cycle plotted in gray.
The vertical dashed blue lines show the eclipse center as determined for Part 1.
The bottom X-axis uses
eclipse center to define orbital phase zero. The upper labels
show the predicted orbital phases with time of ascending node defining zero.
}
\label{fig:lat_like_fold}
\end{figure}

We next investigated the orbital gamma-ray modulation by performing likelihood fits to binary phase-resolved
LAT data. We divided the light curve into 50 phase bins. Our analysis was similar to that 
for the long-term light curve, except that we also held the spectral parameters of \src\ fixed and only 
allowed its flux to vary. The likelihood analysis was performed for the Part 1 and Part 2 light curves
separately, and the resulting light curves and their ratios are shown in Fig. \ref{fig:lat_like_fold}.
The folded light curves are similar in overall properties to the folded probability-weighted
aperture photometry light curves (Fig. \ref{fig:lat_fold}). However, it can now be seen
that the minima during the eclipses are very close to zero flux, and given uncertainties
in the background, are consistent with zero. To characterize these phase-resolved
likelihood light curves we again fitted the same functions that we employed for the
aperture photometry light curves. i.e. an eclipse profile for Part 1, and an eclipse profile
plus sine component for Part 2. We kept the eclipse time parameters (center time, phase duration,
phase of eclipse ingress and egress) held fixed at the values determined from fitting
the aperture photometry light curves and only allowed the fluxes and the phase of the sine wave
to be free. The results are given in Table \ref{table:like_fold_fits} and overplotted
on the light curves in Fig. \ref{fig:lat_fold}.
In this parameterization, the out-of-eclipse fluxes are consistent with being the same for
both Part 1 and Part 2.
The maximum of the sine wave component is again found to be 0.5 in phase away from the eclipse,
i.e. at inferior conjunction.
The sine wave semi-amplitude is 17 $\pm$ 4\% of the out-of-eclipse flux.
{\mybf We note that the difference between the mean out-of-eclipse flux and the flux
at eclipse minimum of (3.1 $\pm$ 0.5) $\times$10$^{-8}$\,\phcm2s significantly exceeds the sine wave
semi-amplitude of (0.70 $\pm$ 0.18) $\times$10$^{-8}$\,\phcm2s.}

\subsection{X-ray Results\label{sect:xray_results}}

\begin{figure}
\includegraphics[width=7.25cm,angle=0]{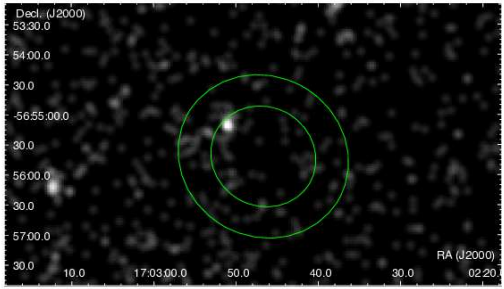}
\caption{
Smoothed Swift XRT image of the region around \src.
LAT 68 and 95\% confidence regions are marked. 
The candidate X-ray counterpart is the brightest source within the 68\% region.
}
\label{fig:xrt_image}
\end{figure}

\begin{figure}
\includegraphics[width=7.25cm,angle=0]{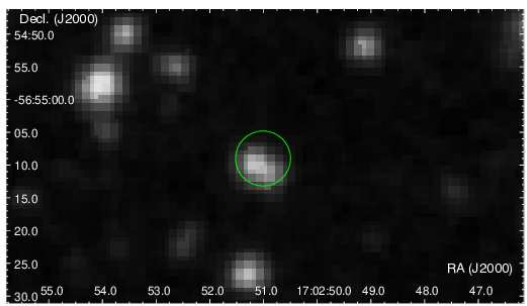}
\caption{
DSS2 red image of the region around the possible XRT counterpart
of \src. The XRT error region is marked.
}
\label{fig:dss2_image}
\end{figure}

A smoothed image obtained from the Swift XRT observations is
shown in Fig. \ref{fig:xrt_image}.
We find that there is an excess within the Fermi 68\% error ellipse
at 17$^h$02$^m$51$^s.01, -56$\degr$55$\arcmin$09\arcsec.$1 with an uncertainty of 4.2\arcsec.
The detection signal-to-noise ratio is 3.7 with 34 counts compared
to an expected background of 10.
Due to the small number of counts it is not possible to obtain
a spectrum of this candidate source. 
Of the total \sqig 17.5\,ks XRT exposure time, only \sqig 3\,ks was obtained during Part 1 of the light curve
which also hampers an investigation of any change in X-ray flux associated with the change
in the gamma-ray orbital modulation.
In Fig. \ref{fig:dss2_image} we show the Deep Sky Survey 2 red image
centered on the XRT location. Within the error region are
several stars which are blended in this image.
Optical spectroscopy of these stars is in process and
will be published later (Swihart et al. in preparation).

\subsection{Radio Results\label{sect:radio_results}}
Although several radio sources were detected within the ATCA field of view, none lie
within the LAT error region. Hence, we are only able to obtain upper limits on
the presence of a radio counterpart. From the summed images at each frequency we obtain
3$\sigma$ upper limits at the source location
of 66, 69, and 69 $\mu$Jy at 2100, 5500 and 9500 MHz respectively.
We note that in the second Fermi LAT catalog of gamma-ray pulsars \citep[``2PC''][]{Abdo2013}, sources
with flux densities $<$ 30\,$\mu$Jy at 1400 MHz are regarded as radio quiet, and that most
pulsars have spectral indices of $-$1.7. Thus, at \sqig2000 MHz radio quiet sources
would have flux densities $\lesssim$ 60\,$\mu$Jy, which is comparable to our upper limits
and hence we cannot yet exclude the presence of a radio pulsar.

\section{Discussion\label{sect:discuss}}

\subsection{The Characteristics of the Orbital Modulation in \src}
The sharp dip seen in both the earlier and later portions of the
light curve is consistent with an eclipse of the gamma-ray emitting region
by a companion star in a highly inclined orbit.
This is thus consistent with the proposed classification
as a redback system.
The presence of strong modulation seen in the power spectrum of the
LAT light curve is exceptional for binary MSPs. The detection of the period is related to the
change in orbital modulation. When tMSPs 
undergo state changes, larger flux variations are seen \citep[e.g.][]{Roy2015} than the
somewhat modest change in orbital profile seen for \src\
which is not accompanied by large changes in the gamma-ray flux
or spectrum.

We next review orbital modulation in other binary MSPs as previously reported from LAT
observations for comparison with \src. For several systems we also consider the
properties of this modulation as found from our program.
The parameters of these systems are summarized in Table \ref{table:other_systems}.
For \src\ we have used the center of the eclipse to define orbital phase 0.
For
most systems discussed here pulse timing orbits have been derived, but no eclipses
have been observed. Hence for these systems typically the time of the ascending node is used to define orbital
phase ($\phi_{AN}$) 0. For a circular orbit where the gamma-ray emission is centered on the neutron star, 
the ascending node will occur 0.25 in phase earlier than an eclipse. i.e. an eclipse would be
expected to occur at superior conjunction of the pulsar at $\phi_{AN}$ = 0.25, and inferior
conjunction of the pulsar (compact object nearest us) would occur at $\phi_{AN}$ = 0.75.

\subsection{Comparison with Orbital Modulation of Gamma-ray Emission in Other Systems}
\subsubsection{Other Eclipsing MSP Systems}

\begin{figure}
\includegraphics[width=7.25cm,angle=0]{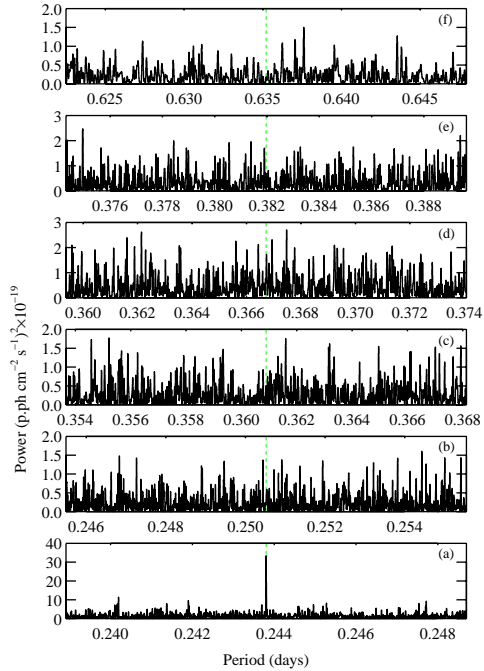}
\caption{
A comparison of the power spectrum of \src\ with the
power spectra of probability-weighted aperture photometry
for likely redback systems for which eclipses have been reported
and are present in the 4FGL DR2 catalog. Power spectra
are centered on the orbital periods (Table \ref{table:other_systems}), which are marked by the vertical
dashed green line.
(a) \src, 
(b) 4FGL\,J1048.6+2340 (PSR J1048+2339),
(c) 4FGL\,J1816.5+4510 (PSR J1816+4510),
(d) 4FGL\,J0427.8$-$6704,
(e) 4FGL\,J1959.5+2048 (PSR B1957+20),
(f) 4FGL\,J2129.8$-$0428 (PSR J2129$-$0429).
}
\label{fig:other_eclipsers}
\end{figure}

To investigate whether any of the other known gamma-ray eclipsing MSP systems 
\citep{Strader2016,Kennedy2020,Clark2021b}
also show any quasi-sinusoidal
modulation, which could be a sign that such modulation depends on the inclination angle of a system,
we calculated the power spectra of the LAT light curves of these sources around their
orbital periods and these are plotted in Fig.  \ref{fig:other_eclipsers}.
For no other eclipsing system is the orbital period detectable in the power spectrum, although
we note that these systems are all fainter than \src\ (Table \ref{table:other_systems}).
For PSR B1957+20, we note that \citet{Wu2012} reported orbital
modulation of $>$ 2.7 GeV gamma rays from Fermi LAT observations with a maximum near the
phase of radio eclipse, i.e. pulsar superior conjunction at $\phi_{AN}$ \sqig 0.25.
{\mybf We also calculated a probability weighted aperture photometry light curve for 4FGL\,J1959.5+2048 (PSR B1957+20)
for energies above 2.7 GeV. However, the power spectrum of this does not show any modulation
at the orbital period, and folding the light curve on the orbital period also did not reveal orbital modulation.}

\subsubsection{4FGL J2039.5-5617 (PSR J2039$-$5617)}

The redback {\mybf system} 4FGL J2039.5-5617 (= PSR J2039$-$5617, 3FGL 2039.6-5618, ``J2039.5'')
was found to have its gamma-ray emission as measured with the LAT modulated on its 0.228 day orbital period with
an approximately sinusoidal profile by \citet{Ng2018}. 
The phasing of the gamma-ray modulation in J2039.5 is that it has a maximum at $\phi_{AN}$ = 0.25 $\pm$ 0.03 \citep{Clark2021a}, i.e. superior conjunction. The maximum of the sinusoidal modulation in \src\ is thus 0.5 out of phase with that in J2039.5, if the eclipse is indeed that of the
compact object in the system.
\citet{Ng2018} suggested that the amplitude of the orbital modulation had decreased after \sqig MJD 57,000.
Such a change in the sinusoidal modulation could indicate a similarity
with \src. 
\citet{Clark2021a} subsequently detected gamma-ray pulsations using the LAT at a period of 2.65 ms.
However, \citet{Clark2021a} found that the orbital modulation increased again after MJD 58,100
and suggested that this could be due to statistical variations rather than intrinsic changes in the source. \citet{Ng2018} and \citet{Clark2021a} noted
that the time of apparent decrease in orbital modulation coincided
with an outburst from the 
blazar candidate 4FGL\,J2052.2$-$5533 (3FGL\,J2051.8$–$5535) which lies
only 1.9\degrees\ from J2039.5 and experienced a flare around
\sqig56,500 to 57,500.

\begin{figure}
\includegraphics[width=7.25cm,angle=0]{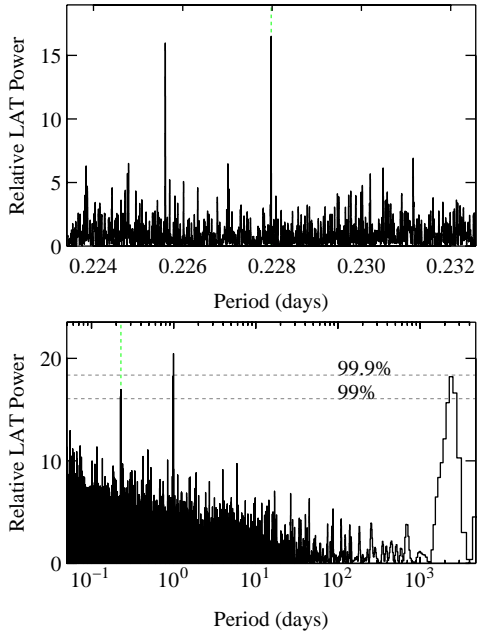}
\caption{
Power spectra of the probability-weighted aperture photometry LAT light curve of 4FGL J2039.5$-$5617 (= PSR J2039$-$5617).
The bottom panel shows a range from 0.05 days to the length of
the light curve, while the top panel shows the detail around the orbital
period of 4FGL J2039.5$-$5617 which is marked with dashed green lines in both panels.
The peak near one day in the bottom panel is a common artifact seen in the power spectra of many LAT sources.
}
\label{fig:2039_power}
\end{figure}

Our power spectrum of the LAT light curve of J2039.5 is shown in
Fig. \ref{fig:2039_power}. In the bottom panel, which covers the entire
frequency range, the strongest peak is near 1 day, 
and this is a commonly seen artifact in the power spectra of LAT
light curves. 
The second strongest peak is at a period of  2429 $\pm$ 112 days and
this long-period/low-frequency peak may be due to contamination from 4FGL\,J2052.2$-$5533.
The third strongest peak is at the orbital period
of J2039.5. The frequency range around this is plotted in
the upper panel of Fig. \ref{fig:2039_power}. 
The period we derive is 0.227978(1), consistent at a 1.8$\sigma$ level 
with the period of 0.227979805(3) days found by \citet{Clark2021a} from pulse timing. 
The relative height of the peak is 17 which corresponds to
an FAP of 0.004 (i.e. 99.6\% significance) for a blind search over the entire frequency range, 
and 4$\times$10$^{-8}$ for a single-frequency trial.
We note the presence of a peak almost as strong as the orbital modulation
close to the orbital period at 0.225601(1) days. While \citet{Clark2021a}
reported variations in the orbital period of J2039.5, these have
a total range of $\Delta P/P$ \sqig10$^{-6}$ and so would not result in the second peak with 
a period difference of \sqig 1\%.
We speculate that this peak is caused by aliasing, although it would be unclear
how this is arising as the frequency difference between these nearby peaks corresponds to a period of \sqig 21.6 days.
We note that J2039.5 is rather fainter than
\src\ which can account for the lower relative height of the modulation in J2039.5
even if both sources had similar variability.

\begin{figure}
\includegraphics[width=7.25cm,angle=0]{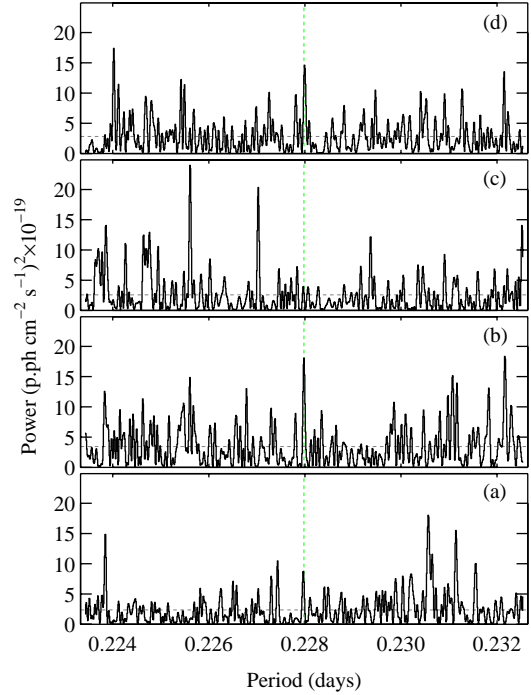}
\caption{
Power spectra of the probability-weighted aperture photometry LAT light curve of 4FGL J2039.5$-$5617 (= PSR J2039$-$5617) divided
into time segments: (a) MJD 55000 - 56000,
(b) MJD 56,000 - 57,000,
(c) MJD 57,000 - 58,000,
(d) MJD 58,000 - 59,000.
The orbital
period of 4FGL J2039.5$-$5617 is marked with dashed vertical green lines in all panels,
and the dashed horizontal gray lines show the mean power levels
}
\label{fig:2039_slice_power}
\end{figure}

We next investigated the rate of change of the orbital peak in the power spectrum 
as a function of light curve length in a similar way to our
analysis of \src. Consistent with \citet{Ng2018} and \citet{Clark2021a},
we find a decrease in the rate of relative peak height change with light curve length between approximately
MJD 57,000 to 58,000. But we again cannot necessarily attribute this to 
a reduction in modulation strength as this coincides with the flare in 4FGL\,J2052.2$-$5533.
To investigate further we calculated power spectra for 1000 day long time intervals
and these are show in Fig. \ref{fig:2039_slice_power}. We find that for segment (c), which covers MJD 57,000 to 58,000,
modulation is not seen at the orbital period, although there is not a large
change in the continuum power level. This may suggest that during this time
the lack of detection of orbital modulation may be because of reduced amplitude, and not just because of
an increase in the background noise level due to the AGN flare.
Nevertheless, the temporal coincidence with this flare does hinder a unambiguous 
determination that there was indeed a reduction in orbital gamma-ray modulation.

\subsubsection{Quasi-Sinusoidal Modulation of Pulsed Emission in 4FGL\,J2339.6$-$0533 (PSR J2339$-$0533)}
The redback PSR J2339$-$0533 (4FGL J2339.6$-$0533, ``J2239.6'')
shows quasi-sinusoidal modulation of the LAT flux on the \sqig0.193 day orbital
period \citep{An2020}. 
{\mybf However,} in this system it is the {\em pulsed} gamma-ray emission of the 2.9 ms pulsar that is 
orbitally modulated
which is difficult to explain. For J2239.6 the orbital maximum
occurs near superior conjunction (when the neutron star is farthest from us) which is similar
to J2039.5, but again apparently 0.5 out of phase with \src. In J2239.6, similar to other binary MSPs where orbital modulation
has been claimed, the gamma-ray modulation was not found from a blind search but relied
on the orbital period already being known.
For J2339.6 we do not detect significant modulation from
our probability-weighted aperture photometry light curve, {\mybf consistent with the report by \citet{An2020}
that orbital modulation is only present in the on-pulse.}

\subsubsection{The Transitional Source XSS J12270−4859 (4FGL\,J1228.0$-$4853)}

XSS J12270−4859 (4FGL\,J1228.0$-$4853, PSR J1227$-$4853) is a transitioning
redback that switched from an LMXB to an MSP state in 2012 \citep{Bassa2014}.
\citet{An2022} reported that the gamma-ray flux {\mybf (60 MeV - 1GeV)} was modulated on
the \sqig0.29 day orbital period of the system, with {\em minimum} at inferior conjunction of the
orbit ($\phi_{AN}$=0.75), and maximum near superior conjunction ($\phi_{AN}$=0.25). 
Curiously, \citet{An2022} found that the orbital gamma-ray
modulation was similar in both the MSP and LMXB states.
However, \citet{Xing2015b} had previously reported the presence of orbital
modulation in LAT gamma-ray observations {\mybf($>$ 200 MeV)} 
with a {\em maximum} near inferior conjunction ($\phi_{AN}$=0.75)
that was only seen after the transition to the MSP state.

\begin{figure}
\includegraphics[width=7.25cm,angle=0]{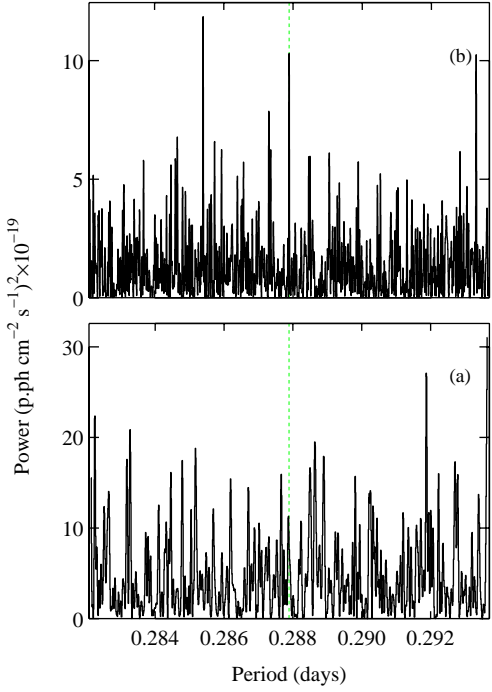}
\caption{
{\mybf
Power spectra of the probability-weighted aperture photometry LAT light curves of the
transitional MSP XSS J12270−4859 (4FGL\,J1228.0$-$4853) for the time intervals of (a) MJD 54,682 - 56,250
and (b) MJD 56,250 - 59,445.
The orbital period (Table \ref{table:other_systems}) is marked by the dashed green lines.
}
}
\label{fig:xss_power}
\end{figure}

\begin{figure}
\includegraphics[width=7.25cm,angle=0]{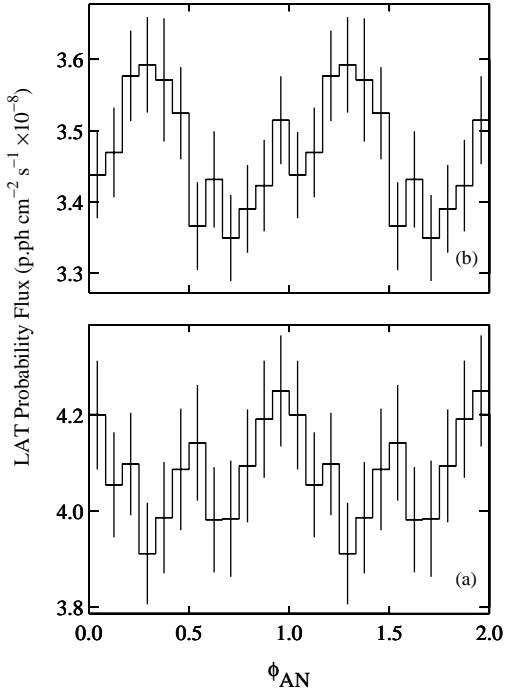}
\caption{
{\mybf
Probability-weighted aperture photometry LAT light curves of the
transitional MSP XSS J12270−4859 (4FGL\,J1228.0$-$4853) folded on the orbital
period for the time ranges of (a) MJD 54,682 - 56,250
and (b) MJD 56,250 - 59,445.
Phase zero was defined as 56700.9070772, corresponding to ``Solution 1'' from \citet{An2022}.
}
}
\label{fig:xss_fold}
\end{figure}

{\mybf
To investigate the presence of orbital modulation in our 100 MeV - 500 GeV light curve of 4FGL\,J1228.0$-$4853 
we calculated power spectra for the time intervals before and after MJD 56,250 (2012 November 19), and these
are shown in Fig. \ref{fig:xss_power}. For the earlier time interval when XSS J12270−4859 was in an LMXB state
we do not detect orbital modulation. However, for the later time interval when the source
had transitioned to an MSP state,we see a peak at a period of 0.287888(3),
consistent with the orbital period. The peak's relative height is 7.8, which corresponds to a single trial FAP of
$4\times10^{-4}$. The detection of orbital modulation only in the MSP state is similar to the result
of \citet{Xing2015b}.
However, the light curves for the same time intervals folded on the orbital period (Fig. \ref{fig:xss_fold})
show that for the later time interval the highest flux is at $\phi_{AN}$\sqig0.25, similar to the result of
\citet{An2022}.
}

\subsubsection{The Black Widows PSR J1311$-$3430 and PSR J2241$−$5236}

Orbital modulation in the {\em off}-pulse emission has also been reported
for the black widow systems PSR J1311$-$3430 \citep[4FGL\,J1311.7$-$3430, ``J1311.7'';][]{Xing2015a,An2017} and 
PSR J2241$−$5236 \citep[4FGL J2241.7$-$5236, ``J2241.7'';][]{An2018}.
For J1311.7 the maximum occurs at $\phi_{AN}$ = 0.8, i.e. near inferior conjunction and minimum at $\phi_{AN}$ =
0.4. For J2241.7 maximum occurs
at superior conjunction ($\phi_{AN}$ 0.25) with a possible secondary maximum at $\phi_{AN}$ = 0.75.

\begin{figure}
\includegraphics[width=7.25cm,angle=0]{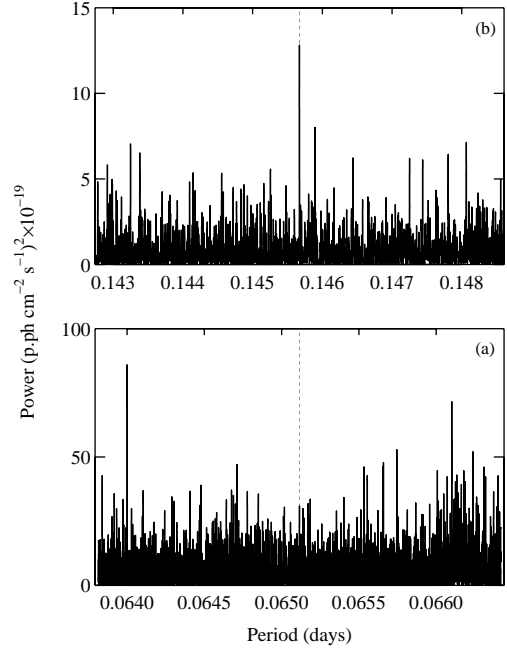}
\caption{
Power spectra of the probability-weighted aperture photometry LAT light curves of the black widow systems 4FGL J1311.7$-$3430 (bottom) and 
4FGL J2241.7$-$5236 (top). The orbital periods (Table \ref{table:other_systems}) are marked by the dashed green lines.
}
\label{fig:bws_power}
\end{figure}

\begin{figure}
\includegraphics[width=7.25cm,angle=0]{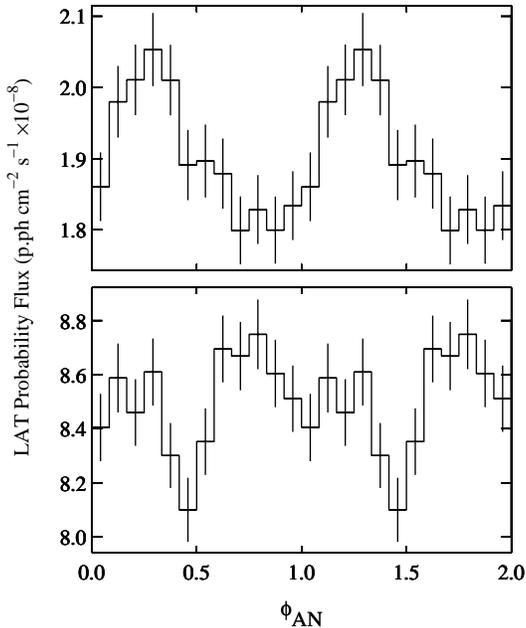}
\caption{
The probability-weighted aperture photometry LAT light curves of the black widow systems 4FGL J1311.7$-$3430 (bottom) and 
4FGL J2241.7$-$5236 (top) folded on their orbital periods (Table \ref{table:other_systems}).
}
\label{fig:bws_fold}
\end{figure}

While these systems did not appear in our blind search for periodicities, power
spectra centered around the known periods do show an indication of modulation for
J2241.7 (Fig. \ref{fig:bws_power}). In the full frequency range blind search
this period is not readily detectable as its relative height is only \sqig13 (FAP = 0.19),
and it is the second highest peak after the 1-day artifact. However, for a single trial
the FAP is \sqig2$\times$10$^{-6}$. No power spectrum peak is seen for J1311.7  (Fig. \ref{fig:bws_power})
- for this source we also investigated using shorter time bins of 100 s due to the short (0.065 day) 
orbital period and no change to the power spectrum was found.
Folding the aperture photometry light curves
on the known periods (Fig. \ref{fig:bws_fold}) does show modulation for both systems. The modulation is thus
detectable in the overall emission from both systems without pulse phase selection.
For both sources we see orbital maximum at the same phases as previously reported \citep{An2017,An2018}. 
There thus appears not to be any large difference in the pulse-averaged profiles 
with the orbital profiles seen in the off-pulse emission, although the amplitudes are lower.

\subsection{Implications for \src}
The gamma-ray variability in \src\ is characterized by two modes of periodic behavior.
In Part 1 of the light curve the only modulation
clearly present is the rather sharp dip that reduces to a level consistent with the background.
Then, this transitions to a combination of a dip in the light curve at a slightly
later phase together with the appearance of
quasi-sinusoidal modulation. This change in periodic behavior is not accompanied
by a significant change in the overall gamma-ray brightness or spectral parameters.
While some aspects of the variability of \src\ have been seen before in other sources, this combination of
eclipses, and transient (on timescales of years) quasi-sinusoidal modulation has not previously been reported.

The light curve dip in Part 2 reaches a similarly low level to that during Part 1,
which implies that the gamma-ray emitting region is still sufficiently small to be totally eclipsed.
However, the minimum occurs later by \sqig0.05 in phase. i.e. the gamma-ray
emission region is trailing the previous center of emission by \sqig 15\degrees.
In addition, the gamma-ray emission now has a non-isotropic component which results
in quasi-sinusoidal orbital modulation with a semi-amplitude of 17 $\pm$ 4\%.
Since pulsations have not yet been detected from \src, we cannot determine whether the
orbital modulation in Part 2 is pulse-phase dependent.

The X-ray brightness of \src\ is low, which suggests that the system has not
transitioned to an accreting LMXB state. 
While we do not yet have an accurate measurement of the X-ray spectrum, or a distance to
the source, we can make a rough estimate of the X-ray to gamma-ray flux ($F_X/F_\gamma$).
At a count rate of \sqig1.4$\times$10$^{-3}$\,cts\,s$^{-1}$, we adopt a power-law spectrum
with a photon-index of 1.75 and N$_{\rm H}$ = 1.25$\times$10$^{21}$\,cm$^{-2}$ \citep{HI4PI2016}, 
using PIMMS\footnote{https://heasarc.gsfc.nasa.gov/cgi-bin/Tools/w3pimms/w3pimms.pl}
the unabsorbed 0.5 - 10 keV flux is 6$\times$10$^{-14}$ \ergscm2s.
The ratio $F_X/F_\gamma$, using the LAT catalog 0.1 - 100 GeV flux (Table \ref{table:other_systems}), is 0.002.
Although this has considerable uncertainty, it is much lower than
the values of \sqig0.25 - 0.4 for tMSPs in the sub-luminous disk state reported
by \citet{Miller2020}. 
{\mybf If the XRT source is not the counterpart of \src, then the ratio of $F_X/F_\gamma$ must be even lower.
Nevertheless, even though \src\ apparently did not transition to an LMXB state, some type of
change did occur to the system.}

Although we do not yet have a Doppler orbit for \src, it appears highly
likely that the ``eclipse'' is indeed the superior conjunction of a pulsar
(non-degenerate object nearest us).
The short ingress and egress, and the minimum being consistent with zero flux, 
indicate an eclipse rather than a change in absorption or emission due to the changing system geometry.  
Were it not an eclipse, the inclination would have to be smaller than about 80\degrees\ given the orbital 
period and typical masses of such systems. The angle dependencies of inverse Compton scattering, absorption due to pair production or Doppler boosted emission would then produce a smoother modulation and would be unlikely to produce zero flux. 
Given the greater penetrating power of gamma-rays compared to X-rays or radio waves,
a much more substantial amount of material is required to cause significant
absorption. Due to this, while radio and X-ray emission can be attenuated
by, for example, winds, to cause a \sqig100\% drop in gamma-rays requires
the body of a star \citep[e.g.][]{Clark2021b}. 
 
For Part 1 of the light curve, where the orbital modulation is apparently less complicated, the total
eclipse duration (ingress start to egress end) is 0.110 $\pm$ 0.038 in phase,
corresponding to a half-angle of 19.8 $\pm$ 6.8\degrees. For comparison, the X-ray eclipse
in 4FGL J0427.8$-$6704 has a half-angle of {\mybf 12.5 $\pm$ 0.3\degrees\ \citep{Strader2016}}.
\citet{Chanan1976} derived eclipse durations for point sources with Roche-lobe filling
companions, although these are degenerate between inclination angle and mass ratio, q
(mass of Roche-lobe filling star/mass of point source).
For an inclination angle of 90\degrees, q $\gtrsim$ 0.25. Similarly, applying
the transit equation of \citet{Seager2003} implies a companion mass $>$ 0.85 \msun\ to avoid it overfilling its Roche lobe.
If, similar to \citet{Strader2016},
we consider the possibility that the system could have an inclination of as low as 75\degrees, the eclipse duration then requires q $\gtrsim$ 1,
although that would imply a surprisingly massive companion star. \citet{Strader2019} find a mean
mass for redback companions of 0.36 $\pm$ 0.04 \msun\ with a $\sigma$ of 0.15 $\pm$ 0.04 \msun, while black widow companions have companions $<$ 0.05 \msun.
If the gamma-ray source is extended, then the constraints on the companion mass are less stringent,
however the sharpness of the ingress and egress (\sqig0.02 in phase) limits the extension of the gamma-ray source to less than \sqig0.2 \rsun\ which sets a lower limit on the companion mass of 0.2 \msun. The companion is very unlikely to be less massive, as this would imply a gamma-ray emission size larger than the size of the star, producing a non-zero residual flux during the eclipse and a slower ingress and egress. 
We conclude that the companion for \src\ is more likely to be similar to those of redback systems than 
the very low mass companions of black widow systems and that, particularly for Part 1, the gamma-ray
emission is not highly extended.

For the spider systems with orbital modulation of gamma-ray flux (excluding eclipses by the secondary)
there is as yet no definite explanation for this.
In addition, there is a difference in systems where the modulation is primarily in the pulsed or non-pulsed emission.
For 4FGL J2039.5-5617, \citet{Ng2018} suggested that the orbital modulation may be due to Compton scattering of soft photons
from the companion and the relativistic pulsar wind.
For 4FGL J2339.6$-$0533, where the {\em pulsed} emission is modulated, \citet{An2020} discuss several scenarios, also including Compton scattering from the companion.
In 4FGL\,J1311.7$-$3430 for the {\em off-}pulse orbital modulation at energies $>$ 200 MeV \citet{An2017} discuss
bulk plasma motion in the IBS.
For systems where quasi-sinusoidal orbital modulation of gamma-ray emission has been reported,
this is generally seen to peak near superior conjunction (Table  \ref{table:other_systems}), unlike what is expected to be the case for \src.
4FGL\,J1311.7$-$3430 does show a peak near inferior conjunction. It also exhibits a secondary peak near superior conjunction.
Such a secondary peak would be more difficult to detect in \src\ due to the eclipse.

If \src\ has not changed between pulsar and accretion regimes, this indicates that some other change has
occurred. For the Part 1 light curve where the only orbital modulation is the relatively sharp eclipse,
we speculate that we are seeing magnetospheric emission from a pulsar in the system that is not
strongly affected by other components in the system.
Then during the latter portion of the light curve (i.e. Part 2) some other change occurred in the system. 
Modulation of the gamma-rays by Compton scattering off the companion star would produce a peak
at superior conjunction. 
{\mybf 
However, modulation due to Doppler boosting of the emission produced in the IBS would produce a maximum at inferior conjunction 
}
if the pulsar wind is collimated away from the star, wrapping around the pulsar as generally observed in redback systems \citep[e.g.][]{An2020}. 
The change in the modulated lightcurve would then be due to variability in the IBS emission region.
The formation and structure of an IBS in a binary MSP depends on the pressure balances
due to winds from the binary components and the pulsar magnetosphere, however
the roles of these are still incompletely understood 
\citep[see e.g.][]{Wadiasingh2017,Wadiasingh2018,vanderMerwe2020}.
{\mybf At least qualitatively, a change in the relative contributions to the gamma-ray emission
from a pulsar and an IBS could explain the observed behavior, including the shift in the timing
of the eclipse and its duration since the IBS emission is not necessarily centered on the pulsar
and has a large spatial extent.}
However, it is challenging to determine why a change to the IBS occurred, why this should be \sqig stable on timescales of years,
and why the flux is not substantially changed due to this transition.

There is, however, some observational evidence that changes to the IBS in binary MSPs may occur.
\citet{Polzin2020} reported that in low-frequency radio observations
of eclipses in the redback system PSR J1816+4510 they found that the radio eclipse mechanism
transitioned between one where pulsar emission was removed from the line of sight
to one where the pulsations were smeared out. These authors attributed this to a tail
of material trailing the pulsar's companion. 
In the redback 47 Tuc W, \citet{Hebbar2021} found that Chandra observations showed
that orbital modulation of X-rays was not always present. They suggested that this
was due to changes in the system's IBS.
For XSS J12270-4859 (4FGL J1228.0−4853) \citet{deMartino2020} found
that the orbital X-ray light curve orbital modulation amplitude varied on timescales
of a few months that could be due to a non-stationary contribution of the IBS.
In addition, optical observations of the redback PSR J1048+2339 (4FGL\,J1048.6+2340)
suggest changes to the IBS on timescales as short as two weeks \citep{Yap2019}.
We speculate that 
the changes in the orbital gamma-ray modulation of \src\ 
might be due to similar changes to an IBS to those that occurred in these systems.
In \src\ we have the advantage of the eclipses which may serve as
a way of more precisely determining emission sites. 
If there is a change in the IBS in \src, then measurements of
the X-ray orbital variability, and particularly any changes in
this associated with a gamma-ray state change, may be a crucial diagnostic.

\section{Conclusion\label{sect:conclude}}

The gamma-ray modulation found in \src\ is exceptional with its combination
of both an eclipse and a quasi-sinusoidal component. The sinusoidal component also has
a maximum near inferior conjunction which is exceptional.
The change in the orbital profile also appears to be unprecedented, which indicates
some change in state, although it is not accompanied by a large change in the gamma-ray flux or spectrum.
To better understand the nature of this source, determining
the optical counterpart and its properties and long-term multiwavelength
monitoring are important to determine what changes occur during the state change
and how this may connect to ``traditional'' tMSPs.
Searches for millisecond pulsations in \src\ are important, and the determination
of Doppler orbits from both pulse timing and optical radial velocity studies are necessary
for determining system parameters, and thus understanding the nature of this system.
Together with the eclipses, which measure the location of emissions region in the system,
these may provide the keys to unlock the physics at work in this system.
Continued monitoring of other sources in the Fermi{\mybf -LAT} catalogs
has the potential to detect a similar system if the same type
of change in its orbital modulation occurs, or the accumulation of additional
data enables the detection of existing modulation. 4FGL\,J2039.5$-$5617 should also
continue to be monitored to determine whether its quasi-sinusoidal modulation does
indeed also change on similar timescales to that in \src.

\acknowledgements

We thank Matthew Kerr and Zorawar Wadiasingh for useful comments.
This work was partially supported by NASA \Fermi\ grant 
80NSSC21K2029 
and also under NASA award number 80GSFC21M0002.
J. Strader acknowledges support from a Packard Fellowship. This work was partially supported by NASA 
grant 80NSSC17K0507 and NSF grant AST-1714825. 
Portions of this work was performed while SJS held a NRC Research Associateship award at the Naval Research Laboratory. Work at the Naval Research Laboratory is supported by NASA DPR S-15633-Y.
The Australia Telescope Compact Array is part of the Australia Telescope National Facility which is 
funded by the Australian Government for operation as a National Facility managed 
by CSIRO who acknowledge the Gomeroi people as the traditional owners of the Observatory site.
We thank the Swift team for undertaking observations and
this work made use of data supplied by the UK Swift Science Data Centre at the University of Leicester.
The \textit{Fermi} LAT Collaboration acknowledges generous ongoing support
from a number of agencies and institutes that have supported both the
development and the operation of the LAT as well as scientific data analysis.
These include the National Aeronautics and Space Administration and the
Department of Energy in the United States, the Commissariat \`a l'Energie Atomique
and the Centre National de la Recherche Scientifique / Institut National de Physique
Nucl\'eaire et de Physique des Particules in France, the Agenzia Spaziale Italiana
and the Istituto Nazionale di Fisica Nucleare in Italy, the Ministry of Education,
Culture, Sports, Science and Technology (MEXT), High Energy Accelerator Research
Organization (KEK) and Japan Aerospace Exploration Agency (JAXA) in Japan, and
the K.~A.~Wallenberg Foundation, the Swedish Research Council and the
Swedish National Space Board in Sweden. Additional support for science analysis during the
operations phase is gratefully acknowledged from the Istituto Nazionale di Astrofisica in
Italy and the Centre National d'\'Etudes Spatiales in France.

\begin{deluxetable}{lcc}
\tablecolumns{8}
\tabletypesize{\small}
\tablewidth{0pc}
\tablecaption{Swift XRT Observations of \src}
\tablehead{
\colhead{Observation Start} & \colhead{Observation End} &\colhead{Exposure} \\
\colhead{(UT Date)}  & \colhead{(UT Date)} &\colhead{(s)} \\
}
\startdata
 2011-11-01 19:58 &	 2011-11-01 21:50 & 1220 \\
 2012-06-26 05:06 &	 2012-06-26 05:12 & 280 \\
 2012-10-30 21:20 &	 2012-10-30 21:36 & 810 \\
 2012-11-06 19:54 &	 2012-11-06 20:16 & 915 \\
\tableline
 2016-11-04 11:51 &	 2016-11-04 19:55 & 1935 \\
 2021-02-06 02:13 &	 2021-02-06 23:20 & 4435 \\
 2021-02-15 01:51 &	 2021-02-16 22:28 & 1145 \\
 2021-03-03 00:00 &	 2021-03-03 05:01 & 1385 \\
 2021-03-04 03:11 &	 2021-03-04 22:28 & 5535 \\
\enddata
\tablecomments{The line indicates the division between Part 1 and Part 2 of the
LAT light curve (MJD 56,345 = 2013-02-22).
}
\label{table:xrt}
\end{deluxetable}

\clearpage

\begin{deluxetable}{lcccc}
\tablecolumns{8}
\tabletypesize{\small}
\tablewidth{0pc}
\tablecaption{Australia Telescope Compact Array Radio Observations of \src}
\tablehead{
\colhead{Observation Start} & \colhead{MJD} &\colhead{Configuration} &\colhead{Center Frequency}  &\colhead{Duration} \\
\colhead{(UT Date)}  & \colhead{} &\colhead{} & \colhead{(MHz)} &\colhead{(minutes)}\\
}
\startdata
2020 Dec 25 &  59208 & 1.5A   &  2100  &             541.1 \\            
2021 Jan 07 &  59221 &  1.5A  &  2100   &            759.3 \\
2021 Jan 14 &  59228 &  EW352 &  2100   &            308.7 \\
2021 Jan 16 &  59230 & EW352  &  2100   &            778.6 \\
2021 Nov 22 &  59540 &  6C    &  5500/9000 &         68.9 \\
\enddata
\tablecomments{The first four observations used 17$^h$02$^m$46$^s.68, 
-56$\degr$55$\arcmin$41\arcsec_.$88 as the target position, while
the last used 17$^h$02$^m$51$^s.45, -56$\degr$55$\arcmin$09\arcsec_.$69.
The ATCA array configurations are the standard names for
the physical locations of the antennas: see
https://www.narrabri.atnf.csiro.au/operations/array\_configurations/configurations.html​
for full details of the antenna spacings in each array configuration.  
}
\label{table:atca}
\end{deluxetable}

\clearpage

\begin{deluxetable}{lcc}
\tablecolumns{8}
\tabletypesize{\small}
\tablewidth{0pc}
\tablecaption{Fits to Periodic Modulation of LAT Light Curve of \src}
\tablehead{
\colhead{Parameter} & \colhead{Part 1} & \colhead{Part 2} 
}
\startdata
Flux Outside Eclipse ($F_{\rm unecl}$) & 4.53 $\pm$ 0.03 & 4.56  $\pm$  0.02 \\
Flux in Eclipse ($F_{\rm ecl}$)& 3.91 $\pm$ 0.22 & 3.85 $\pm$ 0.15 \\
Eclipse Center ({\bfb T$_0$,} MJD) & 57000.168 $\pm$ 0.004 & 57000.181 $\pm$ 0.002 \\
Eclipse Minimum Full Duration ($\Delta$) & 0.044 $\pm$ 0.038 & 0.021 $\pm$ 0.027  \\
Eclipse Ingress Start ($\phi_{\rm ing}$) & 0.940 $\pm$ 0.033 & 0.923 $\pm$ 0.021\\
Eclipse Egress End ($\phi_{\rm egr}$) & 0.050 $\pm$ 0.019  & 0.055 $\pm$ 0.017\\
{\em Eclipse Total Duration (Egress - Ingress)} ($\phi$) &  {\em 0.110 $\pm$ 0.038}   & {\em 0.132 $\pm$ 0.027}  \\
{\em Derived Ingress Duration} ($\phi$) & {\em 0.038 $\pm$ 0.038} &  {\em 0.066 $\pm$ 0.024} \\
{\em Derived Egress Duration} ($\phi$) & {\em 0.028 $\pm$ 0.027}  &  {\em 0.044 $\pm$ 0.022 } \\
\tableline
Sine Wave Half-Amplitude & --- & 0.127 $\pm$ 0.034 \\
Sine Wave Maximum (MJD) & --- & 57000.293 $\pm$ 0.009  \\
{\em Sine Wave Maximum$^1$ ($\phi$)} & --- & {\em 0.51 $\pm$ 0.04} \\
{\em Sine Wave Maximum$^2$ ($\phi$)} & --- & {\em 0.46 $\pm$ 0.04} \\
\tableline
{\bfa $\chi^2_\nu$} & {\bfa 0.78} &  {\bfa 1.04} \\
\tableline
{\em Period (d)} & {\em 0.2438034}  & {\em 0.2438034}  \\
\enddata
\tablecomments{Fits were made to the probability-weighted aperture photometry
light curve with 100 s time bins. Phase 0 corresponds to the center of the full eclipse.
Parameters in italics are derived from the other parameters which were fitted.
The orbital period was held fixed at the value determined from the power spectrum. 
Fluxes are in units of p.ph$\times$10$^{-8}\,$cm$^{-2}$\,s$^{-1}$.
$^1$Relative to the Part 1 eclipse center. $^2$Relative to the Part 2 eclipse center.
Fits are plotted in Fig. \ref{fig:lat_two_fold}.
}
\label{table:lc_fits}
\end{deluxetable}
\clearpage

\begin{deluxetable}{lcc}
\tablecolumns{8}
\tabletypesize{\small}
\tablewidth{0pc}
\tablecaption{Fits to Binary-Phase Resolved LAT Light Curve of \src}
\tablehead{
\colhead{Parameter} & \colhead{Part 1} & \colhead{Part 2} 
}
\startdata
Flux Outside Eclipse ($F_{\rm unecl}$) & 3.92 $\pm$ 0.16 & 4.04  $\pm$  0.12 \\
Flux in Eclipse ($F_{\rm ecl}$)& 0.89 $\pm$ 0.48 & 0.96 $\pm$ 0.44 \\
\tableline
Sine Wave Half-Amplitude & --- & 0.70 $\pm$ 0.18 \\
{\em Sine Wave Maximum (MJD)} & --- & {\em 57000.296 $\pm$ 0.009}  \\
Sine Wave Maximum$^1$ ($\phi$) & --- & 0.52 $\pm$ 0.04 \\
{\em Sine Wave Maximum$^2$ ($\phi$)} & --- & {\em 0.47 $\pm$ 0.04} \\
\tableline
{\bfa $\chi^2_\nu$} & {\bfa 0.58}   & {\bfa 0.997}  \\
\tableline
{\em Period (d)} & {\em 0.2438034}  & {\em 0.2438034}  \\
\enddata
\tablecomments{Fits were made to the phase-resolved fluxes obtained
from likelihood analysis. Eclipse parameters, excluding flux, were held fixed 
at the values obtained from fitting the aperture-photometry light curve given in 
Table \ref{table:lc_fits}. Parameters in italics are derived from the other parameters which were fitted. 
Fluxes are in units of ph$\times$10$^{-8}\,$cm$^{-2}$\,s$^{-1}$.
$^1$Relative to the Part 1 eclipse center. $^2$Relative to the Part 2 eclipse center.
The fits are plotted in Fig. \ref{fig:lat_like_fold}.
}
\label{table:like_fold_fits}
\end{deluxetable}
\clearpage


\begin{deluxetable}{lcccccc}
\tablecolumns{8}
\tabletypesize{\small}
\tablewidth{0pc}
\tablecaption{Selected Binary Millisecond Pulsars}
\tablehead{
\colhead{Name} & \colhead{Photon Flux} & \colhead{Energy Flux}  & \colhead{Orbital Period} & \colhead{Pulse Period} & 
\colhead{Eclipse?} & \colhead{Sine?}\\
\colhead{} & \colhead{(ph cm$^{-2}$\,s$^{-1}$\,$\times10^{-10}$)} & \colhead{(erg\,cm$^{-2}$\,s$^{-1}$\,$\times10^{-12}$)} & 
\colhead{(days)} & \colhead{(ms)} & \colhead{} & \colhead{}
}
\startdata
4FGL\,J1702.7$-$5655 & 31.68 $\pm$  1.10&   29.09$\pm$  1.43 & 0.2438033   & ? & Y & Y/I(?)\\
4FGL\,J1048.6$+$2340 (PSR J1048+2339) &   6.38 $\pm$   0.53 &   5.07 $\pm$ 0.49  & 0.2505191 (a) & 4.67 & Y & N\\
4FGL\,J1816.5$+$4510 (PSR J1816+4510) &  16.91 $\pm$  0.74 & 10.18 $\pm$ 0.52 & 0.3608934817 (b)  & 3.19  & Y & N\\
4FGL\,J0427.8$-$6704 & 6.57 $\pm$  0.43 &   8.57 $\pm$  0.49 & 0.3667200 (c)  & ? & Y & N\\
4FGL\,J1959.5+2048 (PSR B1957+20) & 23.29 $\pm$  1.1 &  16.07 $\pm$  0.89 &  0.38196661 (d) & 1.61 & Y & Y/S \\
4FGL\,J2129.8$-$0428 (PSR J2129-0429) &  11.51$ \pm$  0.72 & 6.68 $\pm$  0.49 & 0.63522741310  (e)  & 7.61  & Y & N\\
\tableline
4FGL\,J1311.7$-$3430 (PSR J1311$-$3430) & 75.68 $\pm$ 1.50 & 60.97 $\pm$ 1.26   & 0.0651157347 (f) & 2.56 & N & Y/I \\
4FGL\,J2241.7$-$5236 (PSR J2241$−$5236) & 46.82 $\pm$ 1.37 & 25.37 $\pm$ 1.11 & 0.1456722372 (g) & 2.19 & N & Y/S \\
4FGL\,J2339.6$-$0533 (PSR J2339$-$0533) & 47.02 $\pm$ 1.26 & 29.16 $\pm$ 0.84 &   0.19309790 (h)  &   2.88 & N & Y/S \\
4FGL\,J2039.5$-$5617 (PSR J2039$-$5617) & 21.83 $\pm$ 0.83 & 15.13 $\pm$ 0.67 & 0.227979805 (i) & 2.65  & N & Y/S \\
4FGL\,J1228.0$-$4853 (PSR J1227$-$4853) & 24.80 $\pm$ 1.08 &  22.58 $\pm$ 1.26 &  0.287887802 (j) & 1.69 & N & Y/S$^\ast$ \\

\enddata
\tablecomments{Photon flux is for the energy range 1 - 100 GeV, energy flux is for the range 100 MeV to 100 GeV.
Both are taken from the 4FGL DR2 catalog (v27).
The ``Eclipse'' column indicates whether an eclipse has been reported in LAT observations. The ``Sine'' column indicates whether quasi-sinusoidal modulation in gamma-rays has been reported, where ``S" and ``I'' indicate orbital maximum is
nearest superior or inferior conjunction of the compact object. 
$^\ast$ For 4FGL\,J1228.0$-$4853 an orbital maximum was reported near inferior conjunction by
\citet{Xing2015b} while \citet{An2022} reported a minimum near that phase from a longer dataset.
References:
(a) \citet{Deneva2016},
(b) \citet{Stovall2014},
(c) \citet{Kennedy2020},
(d) \citet{Arzoumanian1994},
(e) \citet{Kong2018},
(f) \citet{An2017},
(g) \citet{An2018},
(h) \citet{Romani2011},
(i) \citet{Clark2021a},
(j) \citet{deMartino2020}.
The orbital period for \src\ is from this work.
}
\label{table:other_systems}
\end{deluxetable}

\end{document}